\documentstyle[version2,aps]{revtex}
\begin{document}
\draft
\begin{title}
Bogoliubov Quasiparticle Excitations in the Two-Dimensional
$t$$-$$J$ Model
\end{title}
\author{Y. Ohta, T. Shimozato, R. Eder, and S. Maekawa}
\begin{instit}
Department of Applied Physics, Nagoya University,
Nagoya 464-01, Japan
\end{instit}

\begin{abstract}
Using a proposed numerical technique for calculating
anomalous Green's functions characteristic of
superconductivity, we show that the low-lying excitations
in a wide parameter and doping region of the
two-dimensional $t$$-$$J$ model are well described by
the picture of dressed Bogoliubov quasiparticles in the
BCS pairing theory.  The pairing occurs predominantly in
$d_{x^2-y^2}$-wave channel and the energy gap has a size
$\Delta_d$$\simeq$$0.15J$$-$$0.27$$J$ between quarter
and half fillings.  Opening of the superconducting gap
in the photoemission and inverse-photoemission spectrum
is demonstrated.
\end{abstract}


\narrowtext

Recent issues in the theory of high-temperature
superconductivity include the fundamental question
whether the two-dimensional (2D) $t$$-$$J$ model
contains a superconducting phase relevant to cuprate
materials, and if so, what types of superconductivity
the model has.  There are a number of mean-field based,
variational Monte Carlo, and other approximate
calculations, but a commonly agreeable phase diagram
has still been out of our reach.  Thus, at this stage,
unbiased numerically-exact calculations of the model
certainly offer important information on this question.
Among such exact calculations \cite{[1],[2],[3]},
Dagotto and Riera \cite{[1]} have recently found
indications of superconductivity near the region of
phase separation \cite{[4]} by examining equal-time
pairing correlations in finite-size clusters; the
existence of a $d_{x^2-y^2}$-wave condensate has thereby
been conjectured.

In this Letter, we propose a new technique for examining
the low-lying excitation spectrum, i.e., the exact
calculation of anomalous Green's functions for Bogoliubov
quasiparticles in finite-size clusters; we can thereby
calculate the quasiparticle excitation spectrum directly
to examine the superconducting pairing interactions in
the 2D $t$$-$$J$ model.  Then, we show that the low-energy
excitations in a wide parameter ($J/t$) and doping region
of the model are well described by the picture of dressed
Bogoliubov quasiparticles within the BCS
(Bardeen-Cooper-Schrieffer) pairing theory: the singlet
pairing of electrons with opposite momenta occurs near
the Fermi energy unlike in the regime of Bose condensation
of a gas of spin singlets in real space.  The pairing
occurs predominantly in the $d_{x^2-y^2}$-wave channel,
and the gap energy scales with $J$ and has a magnitude
$\Delta_d$$\simeq$$0.15J$$-$$0.27J$.  We also demonstrate
the opening of a superconducting gap in the photoemission
and inverse-photoemission spectrum.  This work provides
the first demonstration of the validity of the BCS pairing
theory for a strongly-correlated electron model.

The $t$$-$$J$ model is defined by the Hamiltonian
\[
H=-t\sum_{<ij>\sigma}
(c^\dagger_{i\sigma}c_{j\sigma}+{\rm H.c.})
+J\sum_{<ij>}({\bf S}_i\cdot {\bf S}_j - {1\over 4}n_i n_j)
\]
where $c^\dagger_{i\sigma}$ ($c_{i\sigma}$) is the
projected electron-creation (annhilation) operator at site
$i$ and spin $\sigma$ ($=\uparrow, \downarrow$) allowing
no doubly occupied sites, ${\bf S}_i$ is the electronic
spin operator, and $n_i$$=n_{i\uparrow}$$+$$n_{i\downarrow}$
is the electron number operator.  The summation is taken
over all the nearest-neighbor pairs $<$$ij$$>$ on the
two-dimensional square lattice.  The average number of
electrons per site, $n$, is defined with the terms half
filling ($n$$=$$1$) and quarter filling ($n$$=$$1/2$).

We adopt the working hypothesis that low-lying states
of the clusters can be described by the microcanonical
version of the BCS pairing theory \cite{[5]}.  Hence,
we assume that a cluster ground-state with an even
electron-number $N$ can be written as
$|\psi^N_0 \rangle$$=$$P_N |{\rm BCS}\rangle$
where $|{\rm BCS}\rangle$ is a BCS wave function
(formulated in terms of `quasiparticles' so that the
effects of strong correlations are already included)
and $P_N$ projects on the $N$-electron subspace.
Similarly, we assume that the low-lying states with
an odd electron-number $N$$+$$1$ can be written as
$|\psi^{N+1}_\nu\rangle$$=$$P_{N+1}
\gamma^\dagger_{{\bf k}\sigma}|{\rm BCS}\rangle$
where $\gamma^\dagger_{{\bf k}\sigma}$ creates a
Bogoliubov quasiparticle with momentum ${\bf k}$.
Then, if we define the following anomalous Green's
functions, our working hypothesis predicts that they
should describe the excitations of Bogoliubov
quasiparticles in the clusters.

The one-particle anomalous Green's function is
defined as
\begin{equation}
G({\bf k},z)=
\langle\psi^{N+2}_0|c^\dagger_{{\bf k}\uparrow}
{1 \over {z-H+E_0}}
c^\dagger_{-{\bf k}\downarrow}|\psi^{N}_0\rangle
\label{eq2}
\end{equation}
where $c^\dagger_{{\bf k}\sigma}$ is the Fourier
transform of $c^\dagger_{i\sigma}$ and $E_0$ is the
ground-state energy.  We define the spectral function
$F({\bf k},\omega)$$=$$-(1/\pi){\rm Im}
G({\bf k},\omega+i\eta)$ with $\eta$$=$$0^+$ and
its frequency integral $F_{\bf k}$$=$$\langle
\psi^{N+2}_0|c^\dagger_{{\bf k}\uparrow}
c^\dagger_{-{\bf k}\downarrow}|\psi^N_0\rangle$.
The hypothesis then predicts
$F({\bf k},\omega)$$=$$F_{\bf k}\delta
(\omega-E_{\bf k})$ with $F_{\bf k}$$=$$z_{\bf k}
\Delta_{\bf k}/2E_{\bf k}$ via the Bogoliubov-Valatin
transformation of operators \cite{[6]}, where $z_{\bf k}$,
$E_{\bf k}$, and $\Delta_{\bf k}$ are the wave-function
renormalization constant, renormalized quasiparticle
energy, and gap function, respectively.  Thus, this
Green's function describes the excitation of one
Bogoliubov quasiparticle in the cluster.  Similarly,
we define the two-particle anomalous Green's function as
\begin{eqnarray}
G({\bf k},{\bf k'},z)&=&
\langle\psi^N_0|
c^\dagger_{-{\bf k'}\downarrow}c_{-{\bf k}\downarrow}
{1\over{z-H+E^N_0}}
c^\dagger_{{\bf k'}\uparrow}c_{{\bf k}\uparrow}
|\psi^N_0\rangle
\nonumber \\
&\;&\;\;\;\;\;
-{n_{-{\bf k}\downarrow}n_{{\bf k}\uparrow} \over {z}}
\delta_{\bf kk'}
\label{eq3}
\end{eqnarray}
with $n_{{\bf k}\sigma}$$=$$\langle\psi^N_0|
c^\dagger_{{\bf k}\sigma} c_{{\bf k}\sigma}|
\psi^N_0\rangle$ and the ground-state energy $E^N_0$.
Only the $N$-particle subspace is involved unlike in
Eq.~(\ref{eq2}).  We define the spectral function
$F({\bf k},{\bf k'},\omega)$ and its frequency integral
$F_{\bf kk'}$ as above.  The hypothesis then predicts
$F({\bf k},{\bf k'},\omega)$$=$$F_{\bf k}
F_{\bf k'}^* \delta(\omega$$-$$E_{\bf k}$$-$$E_{\bf k'})$:
the Green's function describes the excitation of two
Bogoliubov quasiparticles.  Note that Eq.~(\ref{eq3}) may
also be defined with $c^\dagger_{{\bf k'}\sigma}$ and
$c_{{\bf k}\sigma}$ interchanged and with
$n_{{\bf k}\sigma}$ replaced by
$\bar{n}_{{\bf k}\sigma}$$=$$\langle\psi^N_0
|c_{{\bf k}\sigma}c^\dagger_{{\bf k}\sigma}|
\psi^N_0\rangle$; whereas no change occurs in the BCS
theory, some difference appears in the calculated
spectra because of the violation of the anticommutation
relation of the constrained operators; however, we find
the difference to be insignificant.  Thus, by examining
these two anomalous Green's functions, we can see if
the low-energy excitations of the 2D $t$$-$$J$ model
are described by the BCS pairing theory.

The off-diagonal Green's functions Eqs.~(\ref{eq2})
and (\ref{eq3}) may
be evaluated by subtraction of diagonal ones, which in
turn are computed in the standard Lanczos algorithm
\cite{[7]}: e.g., for the one-particle anomalous
Green's function, we prepare the state
$c_{{\bf k}\uparrow}|\psi^{N+2}_0
\rangle$$+$$c^\dagger_{-{\bf k}\downarrow}|
\psi^N_0\rangle$
and calculate the spectral function, from which the
usual single-particle spectral functions for
$c_{{\bf k}\uparrow}|\psi^{N+2}_0\rangle$ and
$c^\dagger_{-{\bf k}\downarrow}|\psi^N_0\rangle$ are
subtracted; thereby $F({\bf k},\omega)$$+$${\rm c.c.}$
is obtained.  The majority of the spectral weight is
cancelled out, remaining only the anomalous part
(compare the spectra shown below).  The imaginary part
of $F({\bf k},\omega)$ is also calculated by using the
state $c_{{\bf k}\uparrow}|\psi^{N+2}_0
\rangle$$+i$$c^\dagger_{-{\bf k}\downarrow}|
\psi^N_0\rangle$ and is always found to vanish.
We use clusters of the size $4\times4$ and
$\sqrt{18}$$\times$$\sqrt{18}$ with periodic boundary
condition; the Green's functions are evaluated for
zero-momentum ground states to simulate the excitations
in the thermodynamic limit.  The Fermi momentum
${\bf k}_{\rm F}$ is located at the $k$-points
where $F({\bf k},\omega)$ and $F({\bf k},{\bf k'},\omega)$
show the lowest-energy peak.

The calculated results for $F({\bf k},\omega)$ are
shown in Fig.~\ref{fig1}.  We choose the average value
of $E_0^N$ and $E_0^{N+2}$ as the value of $E_0$
used in Eq.~(\ref{eq2}).  We find the following, all
of which are consistent with expectations of the BCS
pairing theory:  A pronounced low-energy peak appears
at ${\bf k}_{\rm F}$ and smaller peaks appear at higher
energies for other momenta; the weights of the peaks
are consistent with the BCS form of the condensation
amplitude $F_{\bf k}$ with a maximum at
${\bf k}_{\rm F}$.  The momentum dependence of
$F({\bf k},\omega)$, i.e., the change in sign
under rotation by $\pi/2$ and vanishing weight along
the $k_x$$=$$k_y$ line, is a clear indication of
$d_{x^2-y^2}$-wave pairing.  The size of the energy
gap, which may be estimated directly from the
positions of the peaks, scales well with $J/t$ value.
With decreasing gap size, the peaks at momenta other
than ${\bf k}_{\rm F}$ lose their weight as expected
from the BCS theory.  An important consequence
of $d_{x^2-y^2}$-wave pairing may be seen in the
point-group symmetry of the ground states, i.e., an
alternation of the symmetry between $A_1$ and $B_1$
in, e.g., the 18-site cluster for fillings of 10, 12,
14, and 16 electrons in a fairly wide $J/t$ region.
This alternation, absent in, e.g., the attractive-$U$
Hubbard clusters, suggests the picture that electrons
are added in pairs with $d_{x^2-y^2}$-wave symmetry.
$G({\bf k},z)$ in Eq.~(\ref{eq2}) (and thus
$F({\bf k},\omega)$) reflects the pairing symmetry
via the point-group symmetries of $\psi^N_0$ and
$\psi^{N+2}_0$.  Note that $F({\bf k},{\bf k'},\omega)$
is defined entirely within the $N$-particle subspace
and thus is not affected by this alternation;
nevertheless it shows the same indication of
$d_{x^2-y^2}$-wave pairing as $F({\bf k},\omega)$.

The calculated results for $F({\bf k},{\bf k'},\omega)$
are shown in Figs.~\ref{fig2} and \ref{fig3}.  We
again find that the size of the excitation gap scales
well with $J/t$, and the ${\bf k}$ dependence of the
spectra clearly indicates $d_{x^2-y^2}$-wave pairing.
There are sharp peaks at low energies and broadened
features at higher energies.  As is the case for
$F({\bf k},\omega)$, these high-energy features lose
their weight rapidly with decreasing $J/t$ value,
whereas the peaks at ${\bf k}_{\rm F}$ become sharp
but remain finite with decreasing $J/t$ value.  These
results are consistent with the notion of
weakly-interacting Bogoliubov quasiparticles for
low-lying excitations in the BCS superconductors.

To be more quantitative, let us compare the
calculated spectra $F({\bf k},\omega)$ and
$F({\bf k},{\bf k'},\omega)$ with the BCS predictions
for the spectral functions: the quasiparticle energy
$E_{\bf k}$$=$$\sqrt{\xi_{\bf k}^2+\Delta_{\bf k}^2}$
with one-electron energy $\xi_{\bf k}$$=$$-$$2t_{\rm eff}
(\cos k_x+\cos k_y)$$-$$\mu$ and gap function
$\Delta_{\bf k}=\Delta_d (\cos k_x-\cos k_y)$ is
assumed.  We use the effective hopping parameter
$t_{\rm eff}$ to take into account the quasiparticle
band narrowing.  The chemical potential $\mu$ is chosen
to guarantee vanishing $\xi_{\bf k}$ at
${\bf k}_{\rm F}$.  The value of $\Delta_d$ is then
evaluated by fitting the positions of the low-energy
peaks in $F({\bf k},\omega)$ and
$F({\bf k},{\bf k'},\omega)$.  The renormalization
factor $z_{\rm B}$ (defined by assuming $z_{\bf k}$
to be ${\bf k}$-independent) is estimated from weights
of the low-energy peaks in $F({\bf k},\omega)$ and
$F({\bf k},{\bf k'},\omega)$.  The parameters
$t_{\rm eff}$ and $z_{\rm B}$ reflects the effects of
strong correlation and imply the use of a
{\it dressed\/} Bogoliubov-quasiparticle description.
The fitted quasiparticle spectra are shown by dotted
curves in Figs.~\ref{fig1}--\ref{fig3}.  We find a fair
agreement with the exact spectra, which demonstrates
the validity of the BCS pairing theory for low-lying
excitations in the 2D $t$$-$$J$ model.  This is also
the case at low-doping levels: both $F({\bf k},\omega)$
and $F({\bf k},{\bf k'},\omega)$ exhibit well-defined
low-energy peaks with $d_{x^2-y^2}$-wave symmetry
which can be fitted to the BCS spectra with rather
small $t_{\rm eff}$ and $z_{\rm B}$ values.

The estimated values of the gap parameter $\Delta_d$
are shown in Fig.~\ref{fig4}.  We find that $\Delta_d$
scales well with $J$ and has the value
$\Delta_d$$\simeq$$0.15J$$-$$0.27$$J$ until reaching the
region of phase separation.  The gap value is large
($\,\raise 2pt \hbox{$<\kern-10pt \lower 5pt
\hbox{$\sim$}$}\,$$0.27J$) near half filling and small
($\sim$$0.15J$) around quarter filling.  The
renormalization factor varies over
$z_{\rm B}$$\simeq$$0.2$$-$$1$ with smaller values
at low-doping levels.  Note that the maximum gap-energy
2$\Delta_d/t$$\simeq$$0.9$ ($0.5$) around quarter
filling at $J/t$$=$$2.5$ ($1.5$) (see Fig.~\ref{fig4})
is significantly smaller than the effective
half-bandwidth $4t_{\rm eff}/t$$\simeq$$2.2$ estimated
from the fitting of $F({\bf k},\omega)$ and
$F({\bf k},{\bf k'},\omega)$.  Also, at low-doping
levels, we note that the estimated value
2$\Delta_d$$\simeq$$0.5J$$-$$0.6J$ is smaller than
the effective bandwidth of $\sim$$2J$$-$$4J$.  Thus,
even near phase separation or at low-doping levels,
the superconductivity in the 2D $t$$-$$J$ model is
not in the regime of Bose condensation of a gas of
spin singlets in real space.

The single-particle spectral function $A({\bf k},\omega)$
(see Ref.\cite{[8]} for the definition) simulates
angle-resolved photoemission (PES) and inverse
photoemission (IPES) spectroscopy.  The calculated
spectra are shown in Fig.~\ref{fig5}, which shows
how the superconductivity affects the single-particle
excitations.  We find that with increasing $J/t$ a
gap-like feature appears at ${\bf k}$$=$$(2\pi/3,0)$
which may be interpreted as a superconducting gap as
the BCS dispersion $\pm$$E_{\bf k}$ suggests.  An
associated spectral-weight transfer also indicates
a tendency toward smearing of the jump at
${\bf k}_{\rm F}$ in the momentum distribution
function.  We have also calculated the spin-excitation
spectrum $S({\bf q},\omega)$ (not shown) and compared
it with the bare susceptibility in the BCS mean-field
theory.  We find overall agreement including
spectral-weight distributions and opening of the
superconducting gap: the features are well interpreted
in terms of the particle-hole excitations across the
noninteracting Fermi surface although at
${\bf q}$$=$$(\pi,\pi)$ the effect of nearest-neighbor
antiferromagnetic spin correlations becomes appreciable
in the large-$J/t$ region or at low-doping levels.
Details will be discussed elsewhere.

Summarizing, we have studied the superconductivity
in the 2D $t$$-$$J$ model with the use of a newly
proposed technique for examining the low-lying
excitation spectrum, i.e., the exact calculation of
anomalous Green's functions for Bogoliubov-quasiparticle
excitations characteristic of the superconducting
state.  The validity of the BCS pairing theory for a
strongly-correlated electron model has thereby been
demonstrated for the first time.  We have shown that
the pairing occurs predominantly in the
$d_{x^2-y^2}$-wave channel and the energy gap has
a size $\Delta_d$$\simeq$$0.15J$$-$$0.27J$ in a wide region
between quarter and half fillings.

This work was supported by Priority-Areas Grants from
the Ministry of Education, Science, and Culture of
Japan.  R. E. acknowledges financial support by the
Japan Society for Promotion of Science.  Computations
were partly carried out in the Computer Center of
Institute for Molecular Science, Okazaki National
Research Institutes.

\figure{
Bogoliubov-quasiparticle spectrum
$F({\bf k},\omega)$ for the 16-site cluster with doping
between 4 and 6 holes at (a) $J/t$$=$$0.8$ and
(b) $1.5$, and for the 18-site cluster with doping
between 4 and 6 holes at (c) $J/t$$=$$0.4$ and
(d) $1.5$.  The spectra at other momenta ${\bf k}$
are identical with those shown here except the sign
that follows the $d_{x^2-y^2}$-wave symmetry; the
spectra along the line $k_x$$=$$k_y$ vanish exactly.
${\bf k}_{\rm F}$ is located at $(\pi,0)$ in (a) and
(b), and at $(2\pi/3,0)$ in (c) and (d).  The dotted
curves show the BCS spectral function obtained for
parameter values $\Delta_d/t$$=$$0.13$ in (a) and
$0.29$ in (b) with $t_{\rm eff}/t$$=$$0.55$ and
$z_{\rm B}$$=1.0$, and $\Delta_d/t$$=$$0.05$ in (c)
and $0.27$ in (d) with $t_{\rm eff}/t$$=$$0.45$
and $z_{\rm B}$$=$$0.5$.  We use the value
$\eta/t$$=$$0.15$.
\label{fig1}}

\figure{
Bogoliubov-quasiparticle spectrum
$F({\bf k},{\bf k'},\omega)$ for the 16-site cluster
with 8 holes at $J/t$$=$$0.4$ (left panel) and 2.5
(right panel).  The momentum ${\bf k}$ dependence
is shown with ${\bf k'}$ fixed at $(0,-\pi/2)$.
${\bf k}_{\rm F}$ is located at $(\pi/2,0)$ and
its equivalent points.  The value $\eta/t$$=$$0.15$
is used. The dotted curves show the BCS spectral
function obtained for parameter values
$\Delta_d/t$$=$$0.105$ (left panel) and $0.47$
(right panel) with $t_{\rm eff}/t$$=$$0.55$ and
$z_{\rm B}$$=$$0.7$.
\label{fig2}}

\figure{
As in Fig.~\ref{fig2} but for the 18-site cluster
with 6 holes at $J/t$$=$$0.4$ (left panel) and 1.5
(right panel).  The momentum ${\bf k'}$ is fixed
at $(0,-2\pi/3)$, and ${\bf k}_{\rm F}$ is located
at $(2\pi/3,0)$ and its equivalent points.  The
value $\eta/t$$=$$0.15$ is used.  The BCS spectra
are for the parameter values $\Delta_d/t$$=$$0.05$
(left panel) and 0.285 (right panel) with
$t_{\rm eff}/t$$=$$0.45$ and $z_{\rm B}$$=$$0.5$.
\label{fig3}}

\figure{
Gap parameter $\Delta_d$ for various $J/t$ values
and doping rates. The panel gives the number of
holes (in the 16- or 18-site cluster) in the final
state of the anomalous Green's functions: odd (even)
numbers imply that the results are from the one-particle
(two-particle) anomalous Green's function.  The dashed
lines are a guide to the eye.
\label{fig4}}

\figure{
Single-particle spectral function
$A({\bf k},\omega)$ for the 18-site cluster with
6 holes at $J/t$$=$$0.4$ (left panel) and $1.5$
(right panel).  $\omega$ is measured from the
6-hole ground-state energy, ${\bf k}$'s are
shown in the panels, and the value $\eta$$=$$0.05$
is used.  The thin-dashed curves show the BCS
dispersion $\pm$$E_{\bf k}$ obtained for the parameter
values used in Fig. \ref{fig3}: good agreement is
seen near the Fermi energy while at higher energies
meaningful comparison cannot be made because of the
absence of damping effects.
\label{fig5}}

\end{document}